\documentclass[preprint,showpacs,preprintnumbers,amsmath,amssymb]{revtex4-1}
\usepackage{graphicx}
\usepackage{dcolumn}
\usepackage{bm}
\usepackage{graphicx}
\usepackage{subfigure}
\usepackage{tikz}
\usepackage{siunitx}
\usepackage{float}
\usepackage{booktabs}
\usepackage{url}
\usepackage[utf8]{inputenc}
\usepackage{color}
\usepackage{makecell}
\usepackage{threeparttable}

\usetikzlibrary{shapes,arrows}

\begin{document}

\title{b-baryon semi-tauonic decays in the Standard Model}

\author{Chao Han} \email{hanchao@mail.itp.ac.cn} \quad

\author{Chun Liu} \email{liuc@mail.itp.ac.cn}

\affiliation{CAS Key Lab. of Theor. Phys., Institute of Theoretical 
Physics, Chinese Academy of Sciences, Beijing 100190, China} 

\affiliation{School of Physical Sciences, Univ. of Chinese Academy of 
Sciences, Beijing 100049, China}

\date{\today}

\begin{abstract} 

Within the framework of HQET, 
$\Lambda_{b}\rightarrow\Lambda_{c}\tau\bar{\nu}_{\tau}$ and 
$\Omega_{b}\rightarrow\Omega_{c}^{(*)}\tau\bar{\nu}_{\tau}$ weak decays 
are studied to the order of $1/m_c$ and $1/m_b$.  Helicity amplitudes 
are written down.  Relevant Isgur-Wise functions given by QCD sum rule 
and large $N_c$ methods are applied in the numerical analysis.  The 
baryonic R-ratios $R(\Lambda_c)$ and $R(\Omega_c^{(*)})$ are obtained.   
\end{abstract}


\maketitle

\section{introduction} 

Recent experiments showed that there are some deviation from standard 
theory expectation in B-meson semi-tauonic decays 
\cite{timmurphy.org}.  The ratio 
$R(D^{(*)})\equiv{\rm Br}(B\to D^{(*)}\tau\bar{\nu})/{\rm Br}(B\to D^{(*)} l\bar{\nu})$ 
($l=e,~\mu$) is about 3$\sigma$ larger than the Standard Model (SM) 
prediction.  This attracts a lot of theory attention from new physics 
point of view, for reviews see 
Refs. \cite{Blanke:2019yfv,gori2019tasi,Pich:2019pzg,Li:2018lxi,fajfer2013theory}.  
Nevertheless, a careful SM calculation is needed before any new physics 
conclusion can be drawn.  An alternative place to further check that if 
there is such an anomaly is to look at b-baryon semileptonic decays.  
In this paper, the SM analysis of weak decays 
$\Lambda_{b}\rightarrow\Lambda_{c}\tau \bar{\nu}_{\tau}$ and 
$\Omega_{b}\rightarrow\Omega_{c}^{(*)}\tau\bar{\nu}_{\tau}$ is made in 
detail with tauon mass effects considered.  

Heavy baryons are interesting both experimentally and theoretically for 
their own sake.  They reveal important features of the heavy quark 
physics, and offer a good application ground of QCD.  Especially 
experiments of LHC \cite{aaji2012measurement}, the super B-factory 
\cite{Kou:2018nap} and BES III \cite{Ablikim:2019hff}, \tabularnewline  
as well as previous LEP, LEPII and Tevatron, have been collecting data 
on heavy baryons.  Among various heavy baryons, 
$\Lambda_b$ and $\Omega_b$, as the most basic and simplest ones, 
have been paid much attention.  Their semileptonic weak decays have 
already been analyzed in detail 
\cite{grozin1992sum,dai1996qcd,lee1998analysis,huang2005note,Pervin:2005ve,Ebert:2006rp,Guo:2005qa,Ke:2007tg,Du:2011nj,Jia:2012vt,Du:2013zba,Detmold:2015aaa,Roy:2016txl,Rahmani:2016wgu}.
At the quark-gluon level, the decays are due to weak interaction which 
is described by the universal V-A four-fermion interaction.  At the 
hadron level, semileptonic decay amplitudes are written in terms of 
hadronic matrix elements of weak currents, which are parameterized by 
form factors.  In this paper, helicity amplitude description of 
hadronic transitions by K\"orner {\it et al.} 
\cite{korner1994heavy,kadeer2009helicity,korner1990exclusive,korner1992polarization} 
will be used.  

The form factors of heavy hadron weak transition can be greatly 
simplified in the heavy quark effective theory (HQET) 
\cite{isgur1989weak,georgi1990effective,isgur1991heavy,manohar2007heavy,Neubert:1993mb,Georgi:1990cx}.  
For hadrons containing a single heavy quark, HQET is the right QCD to 
describe them.  It factorizes the perturbatively calculable part out 
from hadronic matrix elements.  Form factors are then described by 
several independent universal form factors which are the so-called 
Isgur-Wise functions.

Nonperturbative methods are needed to calculate Isgur-Wise functions.  
Actually it is at this stage the uncertainty of analytical calculation 
is lack of control.  Among various analytical nonperturbative methods, 
QCD sum rules \cite{shifman1979qcd} and the large $N_c$ limit 
\cite{tHooft:1973alw,witten1979baryons} are outstanding, the former is 
regarded as being rooted in QCD, and the latter is just a limit of QCD.  
They are generally considered to be more close to real QCD.  Both have 
reasonable and consistent ways to estimate uncertainties of calculation.  
Like in \cite{lee1998analysis,Du:2011nj,Du:2013zba}, we will use results 
of both the QCD sum rule and the large $N_c$ methods for these 
Isgur-Wise functions in the final numerical analysis.   Notably 
$\Lambda_b$ semi-tauonic decay was also studied recently 
\cite{gutsche2015semileptonic,shivashankara2015lambda,dutta2016lambda,Li:2016pdv,datta2017phenomenology,azizi2018semileptonic,di2018detailed,bernlochner2018new,bernlochner2018new,bernlochner2019precise,mu2019investigation,bevcirevic2020heavy,rajeev2019implication}.  

The outline of the paper is as follows.  In section 
\uppercase\expandafter{\romannumeral2}, general description of the 
decays in terms of helicity amplitudes is given.  In section 
\uppercase\expandafter{\romannumeral3}, form factors are expressed by 
Isgur-Wise functions which were obtained from the large $N_c$ and QCD 
sum rules.  In section \uppercase\expandafter{\romannumeral4}, numerical 
results are presented.  Section \uppercase\expandafter{\romannumeral5} 
gives the summary.

\section{Form Factors, Helicity Amplitudes and Decay Rates} 

\subsection{Form factors} 

For $\Lambda_{b}\rightarrow\Lambda_{c}$ weak transition, relevant 
baryonic matrix elements are parameterized by form factors 
\begin{equation}
\begin{aligned}
\left\langle\Lambda_{c}\left(p^{\prime},s^{\prime}\right)\left|V^{\mu}\right|\Lambda_{b}(p,s)\right\rangle &=\bar{u}_{\Lambda_{c}}\left(p^{\prime},s^{\prime}\right)\left[f_{1}\gamma^{\mu}+if_{2}\sigma^{\mu\nu}q_{\nu}+f_{3}q^{\mu}\right]u_{\Lambda_{b}}(p,s)\,, \\  
\left\langle\Lambda_{c}\left(p^{\prime},s^{\prime}\right)\left|A^{\mu}\right|\Lambda_{b}(p,s)\right\rangle &=\bar{u}_{\Lambda_{c}}\left(p^{\prime},s^{\prime}\right)\left[g_{1}\gamma^{\mu}+ig_{2}\sigma^{\mu\nu}q_{\nu}+g_{3}q^{\mu}\right]\gamma_{5}u_{\Lambda_{b}}(p,s)\,, 
\end{aligned}
\end{equation}
where $q=p-p^{\prime}$ and 
$\sigma_{\mu\nu}=i\left[\gamma_{u},\gamma_{\nu}\right]/2$, form factors 
$f_i$ and $g_i$ are functions of $q^2$.  It is convenient to reexpress 
the form factors as functions of velocities of baryons, 
\begin{equation}
\begin{aligned}
\left\langle\Lambda_{c}\left(v^{\prime},s^{\prime}\right)\left|V^{\mu}\right|\Lambda_{b}(v,s)\right\rangle &=\bar{u}_{\Lambda_{c}}\left(v^{\prime},s^{\prime}\right)\left(F_{1}(\omega)\gamma^{\mu}+F_{2}(\omega)v^{\mu}+F_{3}(\omega)v^{\prime\mu}\right) u_{\Lambda_{b}}(v,s)\,, \\
\left\langle\Lambda_{c}\left(v^{\prime},s^{\prime}\right)\left|A^{\mu}\right|\Lambda_{b}(v,s)\right\rangle &=\bar{u}_{\Lambda_{c}}\left(v^{\prime},s^{\prime}\right)\left(G_{1}(\omega)\gamma^{\mu}+G_{2}(\omega)v^{\mu}+G_{3}(\omega)v^{\prime\mu}\right)\gamma^{5}u_{\Lambda_{b}}(v,s)\,, 
\end{aligned}
\end{equation}  
where $v$ and $v^{\prime}$ denote four-velocities of $\Lambda_b$ and 
$\Lambda_c$, respectively, $\omega=v\cdot v^{\prime}$, $F_i$ and $G_i$ 
are functions of $\omega$.  
    
Similarly, for the decays of $\Omega_{b} \rightarrow \Omega_{c}^{(*)}$, 
    \begin{equation}
    \begin{aligned}
    \left\langle\Omega_{c}\left(v^{\prime}, s^{\prime}\right)\left|V^{\mu}\right| \Omega_{b}(v)\right\rangle 
    &=\bar{u}_{\Omega_{c}}\left(v^{\prime}, s^{\prime}\right)\left(F_{1}^{\prime} \gamma^{\mu}+F_{2}^{\prime} v^{\mu}+F_{3}^{\prime} v^{\prime \mu}\right) u_{\Omega_{b}}(v, s), 
    \\
    \left\langle\Omega_{c}\left(v^{\prime}, s^{\prime}\right)\left|A^{\mu}\right| \Omega_{b}(v)\right\rangle 
    &=\bar{u}_{\Omega_{c}}\left(v^{\prime}, s^{\prime}\right)\left(G_{1}^{\prime} \gamma^{\mu}+G_{2}^{\prime} v^{\mu}+G_{3}^{\prime} v^{\prime \mu}\right) \gamma^{5} u_{\Omega_{b}}(v, s), 
    \\
    \left\langle\Omega_{c}^{*}\left(v^{\prime}, s^{\prime}\right)\left|V^{\mu}\right| \Omega_{b}(v)\right\rangle &=\bar{u}_{\Omega_{c}^{*} \lambda}\left(v^{\prime}, s^{\prime}\right)\left(N_{1}v^{\lambda} \gamma^{\mu}+N_{2} v^{\lambda} v^{\mu}+N_{3} v^{\lambda} v^{\prime \mu}+N_{4} g^{\lambda \mu}\right) \gamma^{5} u_{\Omega_{b}}(v, s), 
    \\
    \left\langle\Omega_{c}^{*}\left(v^{\prime}, s^{\prime}\right)\left|A^{\mu}\right| \Omega_{b}(v)\right\rangle 
    &=\bar{u}_{\Omega_{c}^{*} \lambda}\left(v^{\prime}, s^{\prime}\right)\left(K_{1} v^{\lambda} \gamma^{\mu}+K_{2} v^{\lambda} v^{\mu}+K_{3} v^{\lambda} v^{\prime \mu}+K_{4} g^{\lambda \mu}\right) u_{\Omega_{b}}(v, s),
    \end{aligned}
    \end{equation}
where the $u_{\Omega_{c}^{*} \lambda}$ is the Rarita-Schwinger spinor for a spin-3/2 particle.

\subsection{Helicity amplitudes}

In analyzing decays, polarization gives detailed physics information.   
The decay $\Lambda_{b}\rightarrow\Lambda_{c}\tau\bar{\nu}_{\tau}$ can be 
thought of as two sub-processes 
$\Lambda_{b}\rightarrow\Lambda_{c}+W_{\text{off-shell}}$ and 
$W_{\text{ off-shell }}\rightarrow\tau+\bar{\nu}_{\tau}$. 
Consider the decay 
$\Lambda_{b}\rightarrow\Lambda_{c}+W_{\text{off-shell}}$ in the rest 
system of $\Lambda_{b}$.  $W_{\text{off-shell}}$ moves in the $+z$ 
direction, and $\Lambda_{c}$ moves in the $-z$ direction.  The momentums 
of $\Lambda_{b}, \Lambda_{c}$ and $W_{\text{off-shell}}$ are 
\begin{equation}
p^\mu=\left(m_{\Lambda_b};0,0,0\right), \quad p^{\prime\mu}=\left(E_{\Lambda_c};0,0,-|\vec{q}|\right), \quad q^{\mu}=\left(q_{0};0,0,|\vec{q}|\right), 
\end{equation}
respectively.  The current is composed of a spin-1 and a spin-0 
components.  The relevant expression of polarization 4-vectors of the 
current is that \cite{auvil1966wave} 
\begin{equation}
\varepsilon^{\mu}(t)=\frac{1}{\sqrt{q^{2}}}\left(q_{0};0,0,|\vec{q}|\right), \quad\varepsilon^{\mu}(\pm 1)=\frac{1}{\sqrt{2}}(0;\mp 1,-i,0), \quad \varepsilon^{\mu}(0)=\frac{1}{\sqrt{q^{2}}}\left(|\vec{q}|;0,0, q_{0}\right)\,.
\end{equation}
The t-label stands for the time-component of the corresponding current.  
Notice that in our case, the tauon mass will be taken into 
consideration, so the time-component of the 
$W_{\mathrm{off}-\text { shell }}$ should be included.  The helicity 
amplitudes are defined in the following, 
    \begin{equation}
    H_{\lambda_{2} \lambda_{W}}^{V, A}=M_{\mu}^{V, A}\left(\lambda_{2}\right) \epsilon^{* \mu}\left(\lambda_{W}\right),    
    \end{equation} 
where $M_{\mu}^{V, A}$ stand for the matrix elements of vector and 
axial vector currents, $\lambda_{2}$ and $\lambda_{W}$ are the 
helicities of the daughter baryon $\Lambda_c$ and the 
$W_{\mathrm{off}-\text{shell}}$, respectively.  Helicity amplitudes are 
then expressed in terms of the form factors. 
    
For $\Lambda_{b}\rightarrow\Lambda_{c}$ transition,  
\begin{equation}\label{1}
\begin{aligned}
\sqrt{q^{2}}H_{\frac{1}{2} t}^{V} 
&=\sqrt{2 m_{\Lambda_{b}} m_{\Lambda_{c}}(1+\omega)}\left(\left(m_{\Lambda_{b}}-m_{\Lambda_{c}}\right) F_1+(m_{\Lambda_{b}}-m_{\Lambda_{c}}\omega)F_2+(-m_{\Lambda_{c}}+m_{\Lambda_{b}} \omega) F_3\right), 
\\
H_{\frac{1}{2} 1}^{V} 
&=-2\sqrt{m_{\Lambda_{b}} m_{\Lambda_{c}}(\omega-1)}F_1,
\\
\sqrt{q^{2}} H_{\frac{1}{2} 0}^{V} 
&=\sqrt{2 m_{\Lambda_{b}} m_{\Lambda_{c}}(\omega-1)}\left(\left(m_{\Lambda_{b}}+m_{\Lambda_{c}}\right) F_1+m_{\Lambda_{c}}(\omega+1)F_2+m_{\Lambda_{b}}(\omega+1)F_3\right), 
\\
\sqrt{q^{2}}H_{\frac{1}{2} t}^{A} 
&=\sqrt{2 m_{\Lambda_{b}} m_{\Lambda_{c}}(\omega-1)}\left(-\left(m_{\Lambda_{b}}+m_{\Lambda_{c}}\right) G_1+(m_{\Lambda_{b}}-m_{\Lambda_{c}}\omega)G_2+(-m_{\Lambda_{c}}+m_{\Lambda_{b}} \omega) G_3\right),  
\\
H_{\frac{1}{2} 1}^{A} 
&=-2 \sqrt{m_{\Lambda_{b}} m_{\Lambda_{c}}(\omega+1)}G_1,
\\
\sqrt{q^{2}} H_{\frac{1}{2} 0}^{A} 
&=\sqrt{2 m_{\Lambda_{b}} m_{\Lambda_{c}}(\omega+1)}\left(\left(m_{\Lambda_{b}}-m_{\Lambda_{c}}\right) G_1-m_{\Lambda_{c}}(\omega-1)G_2-m_{\Lambda_{b}}(\omega-1)G_3\right). 
\end{aligned}
\end{equation}

Similarly for $\Omega_{b}\rightarrow\Omega_{c}$ transition, 

\begin{equation}\label{2}
\begin{aligned}
\sqrt{q^{\prime2}}H_{\frac{1}{2} t}^{\prime V} 
&=\sqrt{2 m_{\Omega_{b}} m_{\Omega_{c}}(1+\omega)}\left(\left(m_{\Omega_{b}}-m_{\Omega_{c}}\right) F^{\prime}_1+(m_{\Omega_{b}}-m_{\Omega_{c}}\omega)F^{\prime}_2+(-m_{\Omega_{c}}+m_{\Omega_{b}} \omega) F^{\prime}_3\right), 
\\
H_{\frac{1}{2} 1}^{\prime V} 
&=-2\sqrt{m_{\Omega_{b}} m_{\Omega_{c}}(\omega-1)}F^{\prime}_1,
\\
\sqrt{q^{\prime2}} H_{\frac{1}{2} 0}^{\prime V} 
&=\sqrt{2 m_{\Omega_{b}} m_{\Omega_{c}}(\omega-1)}\left(\left(m_{\Omega_{b}}+m_{\Omega_{c}}\right) F^{\prime}_1+m_{\Omega_{c}}(\omega+1)F^{\prime}_2+m_{\Omega_{b}}(\omega+1)F^{\prime}_3\right), 
\\
\sqrt{q^{\prime2}}H_{\frac{1}{2} t}^{\prime A} 
&=\sqrt{2 m_{\Omega_{b}} m_{\Omega_{c}}(\omega-1)}\left(-\left(m_{\Omega_{b}}+m_{\Omega_{c}}\right) G^{\prime}_1+(m_{\Omega_{b}}-m_{\Omega_{c}}\omega)G^{\prime}_2+(-m_{\Omega_{c}}+m_{\Omega_{b}} \omega) G^{\prime}_3\right),  
\\
H_{\frac{1}{2} 1}^{\prime A} 
&=-2 \sqrt{m_{\Omega_{b}} m_{\Omega_{c}}(\omega+1)}G^{\prime}_1,
\\
\sqrt{q^{\prime2}} H_{\frac{1}{2} 0}^{\prime A} 
&=\sqrt{2 m_{\Omega_{b}} m_{\Omega_{c}}(\omega+1)}\left(\left(m_{\Omega_{b}}-m_{\Omega_{c}}\right) G^{\prime}_1-m_{\Omega_{c}}(\omega-1)G^{\prime}_2-m_{\Omega_{b}}(\omega-1)G^{\prime}_3\right), 
\end{aligned}
\end{equation}
and for $\Omega_{b}\rightarrow\Omega_{c}^{*}$ transition, 
\begin{equation}\label{3}
\begin{aligned}
\sqrt{q^{\prime\prime 2}} H_{\frac{1}{2} 0}^{\prime\prime V}
&=\sqrt{\frac{2}{3}}
(\omega+1)\sqrt{2m_{\Omega_{b}}m_{\Omega_{c}^{*}}(\omega-1)}(m_{\Omega_{b}}-m_{\Omega_{c}^{*}})N_1
\\
&-
\sqrt{\frac{2}{3}}
(\omega^2-1)\sqrt{2m_{\Omega_{b}}m_{\Omega_{c}^{*}}(\omega-1)}m_{\Omega_{c}^{*}}N_2
\\
&-
\sqrt{\frac{2}{3}}
(\omega^2-1)\sqrt{2m_{\Omega_{b}}m_{\Omega_{c}^{*}}(\omega-1)}m_{\Omega_{b}}N_3
\\
&-
\sqrt{\frac{2}{3}} \sqrt{2m_{\Omega_{b}}m_{\Omega_{c}^{*}}(\omega-1)}(m_{\Omega_{b}}\omega-m_{\Omega_{c}^{*}})N_4,
\\
\sqrt{q^{\prime\prime 2}} H_{\frac{1}{2} 0}^{\prime\prime A}
&=\sqrt{\frac{2}{3}}
(\omega-1)\sqrt{2m_{\Omega_{b}}m_{\Omega_{c}^{*}}(\omega+1)}(m_{\Omega_{b}}+m_{\Omega_{c}^{*}})K_1
\\
&+
\sqrt{\frac{2}{3}}
(\omega^2-1)\sqrt{2m_{\Omega_{b}}m_{\Omega_{c}^{*}}(\omega+1)}m_{\Omega_{c}^{*}}K_2
\\
&+
\sqrt{\frac{2}{3}}
(\omega^2-1)\sqrt{2m_{\Omega_{b}}m_{\Omega_{c}^{*}}(\omega+1)}m_{\Omega_{b}}K_3
\\
&+
\sqrt{\frac{2}{3}} \sqrt{2m_{\Omega_{b}}m_{\Omega_{c}^{*}}(\omega+1)}(m_{\Omega_{b}}\omega-m_{\Omega_{c}^{*}})K_4,
\\
\sqrt{q^{\prime\prime 2}} H_{\frac{1}{2} t}^{\prime\prime V}
&=\sqrt{\frac{2}{3}}
(\omega+1)\sqrt{2m_{\Omega_{b}}m_{\Omega_{c}^{*}}(\omega-1)}(m_{\Omega_{b}}+m_{\Omega_{c}^{*}})N_1
\\
&-
\sqrt{\frac{2}{3}}
(\omega-1)\sqrt{2m_{\Omega_{b}}m_{\Omega_{c}^{*}}(\omega+1)}(m_{\Omega_{b}}-m_{\Omega_{c}^{*}}\omega)N_2
\\
&-
\sqrt{\frac{2}{3}}
(\omega-1)\sqrt{2m_{\Omega_{b}}m_{\Omega_{c}^{*}}(\omega+1)}(m_{\Omega_{b}}\omega-m_{\Omega_{c}^{*}})N_3
\\
&-
\sqrt{\frac{2}{3}}
(\omega-1)\sqrt{2m_{\Omega_{b}}m_{\Omega_{c}^{*}}(\omega+1)}m_{\Omega_{b}}N_4,
\\
\sqrt{q^{\prime\prime 2}} H_{\frac{1}{2} t}^{\prime\prime A}
&=\sqrt{\frac{2}{3}}
(\omega+1)\sqrt{2m_{\Omega_{b}}m_{\Omega_{c}^{*}}(\omega-1)}(m_{\Omega_{b}}-m_{\Omega_{c}^{*}})K_1
\\
&+
\sqrt{\frac{2}{3}}
(\omega+1)\sqrt{2m_{\Omega_{b}}m_{\Omega_{c}^{*}}(\omega-1)}(m_{\Omega_{b}}-m_{\Omega_{c}^{*}}\omega)K_2
\\
&+
\sqrt{\frac{2}{3}}
(\omega+1)\sqrt{2m_{\Omega_{b}}m_{\Omega_{c}^{*}}(\omega-1)}(m_{\Omega_{b}}\omega-m_{\Omega_{c}^{*}})K_3
\\
&+
\sqrt{\frac{2}{3}} (\omega+1)\sqrt{2m_{\Omega_{b}}m_{\Omega_{c}^{*}}(\omega-1)}m_{\Omega_{b}}K_4,
\\
H_{\frac{1}{2} 1}^{\prime\prime V}
&
=
-\sqrt{\frac{1}{3}} \sqrt{2m_{\Omega_{b}}m_{\Omega_{c}^{*}}}2(\omega+1)N_{1}+
\sqrt{\frac{1}{3}} \sqrt{2m_{\Omega_{b}}m_{\Omega_{c}^{*}}(\omega-1)}N_{4},
\\
{H_{\frac{3}{2} 1}^{\prime\prime V}}  &=-\sqrt{2m_{\Omega_{b}}m_{\Omega_{c}^{*}}(\omega-1)}N_{4},
\\
H_{\frac{1}{2} 1}^{\prime\prime A}
&=
-\sqrt{\frac{1}{3}}\sqrt{2m_{\Omega_{b}}m_{\Omega_{c}^{*}}(\omega+1)}2(\omega-1)K_{1}+\sqrt{\frac{1}{3}} \sqrt{2m_{\Omega_{b}}m_{\Omega_{c}^{*}}(\omega+1)}K_{4},
\\
{H_{\frac{3}{2} 1}^{\prime\prime A}}  &=\sqrt{2m_{\Omega_{b}}m_{\Omega_{c}^{*}}(\omega+1)} K_{4}.
\end{aligned}
\end{equation}
Other relations can be obtained by relations: 
$H_{-\lambda_{2},-\lambda_{W}}^{V}=H_{\lambda_{2},\lambda_{W}}^{V},~ H_{-\lambda_{2},-\lambda_{W}}^{A}=-H_{\lambda_{2},\lambda_{W}}^{A}$.  
At last, the following relation is needed, 
$H_{\lambda_{2}\lambda_{W}}=H_{\lambda_{2}\lambda_{W}}^{V}-H_{\lambda_{2}\lambda_{W}}^{A}$.  
Decay rates can be given in terms of these helicity amplitudes.   

\subsection{Decay rates}

The differential decay rate $\mathrm{d}\Gamma/\mathrm{d}\omega$ is 
obtained as following \cite{kadeer2009helicity,korner1990exclusive}, 
\begin{equation}
\begin{aligned} 
\frac{\mathrm{d} \Gamma(\Lambda_b\rightarrow\Lambda_c\tau\bar{\nu}_\tau)}{\mathrm{d} \omega}
=& 
\frac{G_{F}^{2}}{(2 \pi)^{3}}
\left|V_{c b}\right|^{2} 
\frac{\left(q^{2}-m_{\tau}^{2}\right)^{2} m_{\Lambda_{c}}^{2} \sqrt{\omega^2-1}}{12 m_{\Lambda_{b}} q^{2}} 
\\
& 
\times
[
\left(1+\dfrac{m_{\tau}^2}{2 q^2}\right)\left|H_{\frac{1}{2} 1}\right|^{2}
+\left(1+\dfrac{m_{\tau}^2}{2 q^2}\right)\left|H_{-\frac{1}{2}-1}\right|^{2}
+\left(1+\dfrac{m_{\tau}^2}{2 q^2}\right)\left|H_{\frac{1}{2} 0}\right|^{2}
\\
&
+\left(1+\dfrac{m_{\tau}^2}{2 q^2}\right)\left|H_{-\frac{1}{2} 0}\right|^{2}
+\frac{3 m_{\tau}^2}{2 q^{2}} \left|H_{\frac{1}{2} t}\right|^{2}
+\frac{3 m_{\tau}^2}{2 q^{2}} \left|H_{-\frac{1}{2} t}\right|^{2}]
\\
=&
\frac{\mathrm{d} \Gamma_{T_{+}}}{\mathrm{d} \omega}
+\frac{\mathrm{d} \Gamma_{T_{-}}}{\mathrm{d} \omega}
+\frac{\mathrm{d} \Gamma_{L_{+}}}{\mathrm{d} \omega}
+\frac{\mathrm{d} \Gamma_{L_{-}}}{\mathrm{d} \omega}
+\frac{\mathrm{d} \Gamma_{t_{+}}}{\mathrm{d} \omega}
+\frac{\mathrm{d} \Gamma_{t_{-}}}{\mathrm{d} \omega},
\end{aligned}\label{lambda}
\end{equation}
where $G_F$ is the Fermi coupling constant, $V_{cb}$ is the CKM matrix 
element, and 
$\displaystyle\frac{\mathrm{d}\Gamma_{T_{\pm}}}{\mathrm{d}\omega}, \frac{\mathrm{d}\Gamma_{L_{\pm}}}{\mathrm{d}\omega}$, 
and $\displaystyle\frac{\mathrm{d}\Gamma_{t_{\pm}}}{\mathrm{d}\omega}$ 
are defined as the transverse, longitudinal and time-component 
contribution to the decay rate with $\pm$ denoting the final baryon 
helicity.

Following the same method, we get that for 
$\Omega_{b}\rightarrow \Omega_{c}\tau \overline{\nu}_{\tau}$,  
\begin{equation}\label{omega}
\begin{aligned} 
\frac{\mathrm{d} \Gamma(\Omega_{b}\rightarrow \Omega_{c}\tau \overline{\nu}_{\tau})}{\mathrm{d} \omega}=
& 
\frac{G_{F}^{2}}{(2 \pi)^{3}}
\left|V_{c b}\right|^{2} 
\frac{\left(q^{2}-m_{\tau}^{2}\right)^{2} m_{\Omega_{c}}^{2} \sqrt{\omega^2-1}}{12 m_{\Omega_{b}} q^{2}} 
\\
& 
\times
[
\left(1+\dfrac{m_{\tau}^2}{2 q^2}\right)\left|H^{\prime}_{\frac{1}{2} 1}\right|^{2}
+\left(1+\dfrac{m_{\tau}^2}{2 q^2}\right)\left|H^{\prime}_{-\frac{1}{2}-1}\right|^{2}
+\left(1+\dfrac{m_{\tau}^2}{2 q^2}\right)\left|H^{\prime}_{\frac{1}{2} 0}\right|^{2}
\\
&
+\left(1+\dfrac{m_{\tau}^2}{2 q^2}\right)\left|H^{\prime}_{-\frac{1}{2} 0}\right|^{2}
+\frac{m_{\tau}^{2}}{2 q^{2}} 3\left(\left|H^{\prime}_{\frac{1}{2} t}\right|^{2}+\left|H^{\prime}_{-\frac{1}{2} t}\right|^{2}\right)]
\\
&
=\frac{\mathrm{d} \Gamma_{T^{\prime}_{+}}}{\mathrm{d} \omega}
+\frac{\mathrm{d} \Gamma_{T^{\prime}_{-}}}{\mathrm{d} \omega}
+\frac{\mathrm{d} \Gamma_{L^{\prime}_{+}}}{\mathrm{d} \omega}
+\frac{\mathrm{d} \Gamma_{L^{\prime}_{-}}}{\mathrm{d} \omega}
+\frac{\mathrm{d} \Gamma_{t^{\prime}_{+}}}{\mathrm{d} \omega}
+\frac{\mathrm{d} \Gamma_{t^{\prime}_{-}}}{\mathrm{d} \omega},
\end{aligned}
\end{equation}
\\
and for 
$\Omega_{b} \rightarrow \Omega_{c}^{*} \tau \overline{\nu}_{\tau}$, 
\begin{equation}\label{omega*}
\begin{aligned} 
\frac{\mathrm{d} \Gamma(\Omega_{b} \rightarrow \Omega_{c}^{*} \tau \overline{\nu}_{\tau})}{\mathrm{d} \omega}
&
=\frac{G_{F}^{2}}{(2 \pi)^{3}}
\left|V_{c b}\right|^{2} 
\frac{\left(q^{2}-m_{\tau}^{2}\right)^{2} m_{\Omega^{*}_{c}}^{2} \sqrt{\omega^2-1}}{12 m_{\Omega_{b}} q^{2}} 
\\
& 
\times
[\left(1+\dfrac{m_{\tau}^2}{2 q^2}\right)\left|H^{\prime\prime}_{\frac{3}{2} 1}\right|^{2}
+\left(1+\dfrac{m_{\tau}^2}{2 q^2}\right)\left|H^{\prime\prime}_{-\frac{3}{2}-1}\right|^{2}
+\left(1+\dfrac{m_{\tau}^2}{2 q^2}\right)\left|H^{\prime\prime}_{\frac{1}{2} 1}\right|^{2}
\\
&
+\left(1+\dfrac{m_{\tau}^2}{2 q^2}\right)\left|H^{\prime\prime}_{-\frac{1}{2}-1}\right|^{2} 
+\left(1+\dfrac{m_{\tau}^2}{2 q^2}\right)\left|H^{\prime\prime}_{\frac{1}{2} 0}\right|^{2}
+\left(1+\dfrac{m_{\tau}^2}{2 q^2}\right)\left|H^{\prime\prime}_{-\frac{1}{2} 0}\right|^{2}
\\
&
+
\frac{m_{\tau}^{2}}{2 q^{2}} 3\left(\left|H^{\prime\prime}_{\frac{1}{2} t}\right|^{2}
+\left|H^{\prime\prime}_{-\frac{1}{2} t}\right|^{2}\right) ]
\\
&
=\frac{\mathrm{d} \Gamma^{\prime\prime}_{T_{2+}}}{\mathrm{d} \omega}
+\frac{\mathrm{d} \Gamma^{\prime\prime}_{T_{2-}}}{\mathrm{d} \omega}
+\frac{\mathrm{d} \Gamma^{\prime\prime}_{T_{1+}}}{\mathrm{d} \omega}
+\frac{\mathrm{d} \Gamma^{\prime\prime}_{T_{1-}}}{\mathrm{d} \omega}
+\frac{\mathrm{d} \Gamma^{\prime\prime}_{L_{+}}}{\mathrm{d} \omega}
+\frac{\mathrm{d} \Gamma^{\prime\prime}_{L_{-}}}{\mathrm{d} \omega}
+\frac{\mathrm{d} \Gamma^{\prime\prime}_{t_{+}}}{\mathrm{d} \omega}
+\frac{\mathrm{d} \Gamma^{\prime\prime}_{t_{-}}}{\mathrm{d} \omega},
\end{aligned}
\end{equation}
where 
$\displaystyle\frac{\mathrm{d} \Gamma^{\prime\prime}_{T_{2\pm}}}{\mathrm{d}\omega}$ 
and $\displaystyle\frac{\mathrm{d}\Gamma^{\prime\prime}_{T_{1\pm}}}{\mathrm{d}\omega}$ 
correspond to $H^{\prime\prime}_{\pm\frac{3}{2} \pm1}$ and 
$H^{\prime\prime}_{\pm\frac{1}{2} \pm1}$, respectively.  

\section{HQET with QCD Sum rule and Large $N_c$}

\subsection{HQET}

The form factors in HQET can be simplified in terms of Isgur-Wise functions. For $\Lambda_b\to\Lambda_c$ 
at the leading order of heavy quark expansion, there is only one 
Isgur-Wise function $\xi(\omega)$ 
\cite{isgur1991heavy,manohar2007heavy,Neubert:1993mb}, 
\begin{equation} 
 \left\langle\Lambda_{c}\left(v^{\prime},s^{\prime}\right)\left|\bar{h}_{v^{\prime}}^{(c)}\Gamma h_{v}^{(b)}\right|\Lambda_{b}(v, s)\right\rangle=\xi(\omega) \bar{u}_{\Lambda_{c}}\left(v^{\prime}, s^{\prime}\right) \Gamma u_{\Lambda_{b}}(v, s),   
\end{equation} 
where $h_{v}^{(Q)}$ denotes the heavy quark field defined in the HQET 
with velocity $v$, and $\Gamma$ stands for any gamma matrices.  
$\xi(\omega)$ is normalized at the zero recoil, $\xi(1)=1$.  

When $1/m_Q$ correction is taken into consideration, another Isgur-Wise 
function $\chi$ and a mass parameter $\bar{\Lambda}$ appear.  The 
subleading Isgur-Wise function $\chi(\omega)$ is defined by
\begin{equation}
\begin{aligned} &\left\langle\Lambda_{c}\left(v^{\prime}\right)\left|\mathrm{T} \bar{h}_{v^{\prime}}^{(c)} \Gamma h_{v}^{(b)} i \int d^{4} x \frac{1}{2 m_{Q}} \bar{h}_{v}^{(Q)}(x)(i D)^{2} h_{v}^{(Q)}(x)\right| \Lambda_{b}(v)\right\rangle 
=& \frac{\bar{\Lambda}}{m_{Q}} \chi(\omega) \bar{u}_{\Lambda_{c}}\left(v^{\prime}\right) \Gamma u_{\Lambda_{b}}(v) \end{aligned}\,,
\end{equation}
where $\bar{\Lambda}$ is the heavy baryon mass in HQET,  
$\bar{\Lambda}=m_{\Lambda_{Q}}-m_{Q}$.  

Including $\alpha_s$ and $\Lambda_{\mathrm{QCD}}/m_{c,b}$ corrections, 
the form factors are given as following 
\cite{isgur1991heavy,manohar2007heavy,Neubert:1993mb}, 
\begin{equation}\label{ff1}
\begin{aligned}
F_{1}
&=
C(\mu) \xi(\omega)+
C(\mu) \left(\frac{\bar{\Lambda}}{2 m_{c}}+\frac{\bar{\Lambda}}{2 m_{b}}\right)
[2 \chi(\omega)+\xi(\omega)],
\\
G_{1}
&=C(\mu) \xi(\omega)+
C(\mu) \left(\frac{\bar{\Lambda}}{2 m_{c}}+\frac{\bar{\Lambda}}{2 m_{b}}\right)\left[2 \chi(\omega)+\frac{\omega-1}{\omega+1} \xi(\omega)\right],
\\
F_{2}&=G_{2}=-C(\mu)\frac{\bar{\Lambda}}{m_{c}(\omega+1)} \xi(\omega),
\\
F_{3}&=-G_{3}=-C(\mu)\frac{\bar{\Lambda}}{m_{b}(\omega+1)} \xi(\omega),
\end{aligned}
\end{equation}
where the perturbative QCD coefficient in the leading logarithmic 
approximation is 
\begin{equation}
C(\mu)=\left[\frac{\alpha_{s}\left(m_{b}\right)}{\alpha_{s}\left(m_{c}\right)}\right]^{-6 / 25}\left[\frac{\alpha_{s}\left(m_{c}\right)}{\alpha_{s}(\mu)}\right]^{a_{L}(\omega)},    
\end{equation}
and $a_{L}(\omega)=\displaystyle\frac{8}{27}[\omega r(\omega)-1]$,
$r(\omega)=\displaystyle\frac{1}{\sqrt{\omega^{2}-1}}\ln\left(\omega+\sqrt{\omega^{2}-1}\right)$.  

For  $\Omega_{b(c)}^{(*)}$ cases, similarly, based on the standard 
tensor method \cite{isgur1991heavy,Georgi:1990cx}, we denote 
$\Omega_Q$ and $\Omega_Q^{*}$ as $B_{\mu}^{1}$ and $B_{\mu}^{2}$ 
respectively, 
\begin{equation}
 B_{\mu}^{1}(v, s)=\frac{1}{\sqrt{3}}\left(\gamma_{\mu}+v_{\mu}\right) \gamma^{5} u(v, s), \qquad B_{\mu}^{2}(v, s)=u_{\mu}(v, s).  
\end{equation}
In the leading order of heavy quark expansion, the fourteen form factors 
are reduced to two Isgur-Wise functions which are defined as, 
\begin{equation}
{\left\langle\Omega_{c}^{M}\left|\bar{h}_{v^{\prime}}^{(c)} \Gamma h_{v}^{(b)}\right| \Omega_{b}^{N}\right\rangle= C \bar{B}_{\mu}^{M} \Gamma B^{N}\left[-g^{\mu \nu} \xi_{1}(\omega)+v^{\mu} v^{\prime \nu} \xi_{2}(\omega)\right]}.     
\end{equation}
The form factors are expressed as \cite{Boyd:1990xu}, 
\begin{equation}\label{ff2}
\begin{array}{l}
{F_{1}^{\prime}=\displaystyle\frac{-\omega}{3} C(\mu)\xi_{1}+\frac{\omega^{2}-1}{3} C(\mu)\xi_{2}, \qquad G_{1}^{\prime}=\frac{-\omega}{3} C(\mu)\xi_{1}+\frac{\omega^{2}-1}{3} C(\mu)\xi_{2}}, \\
{F_{2}^{\prime}=\displaystyle\frac{2}{3} C(\mu)\xi_{1}+\frac{2(1-\omega)}{3} C(\mu)\xi_{2}, \qquad G_{2}^{\prime}=\frac{2}{3} C(\mu)\xi_{1}+\frac{-2(1+\omega)}{3} C(\mu)\xi_{2}}, \\ 
{F_{3}^{\prime}=\displaystyle\frac{2}{3} C(\mu)\xi_{1}+\frac{2(1-\omega)}{3} C(\mu)\xi_{2}, \qquad G_{3}^{\prime}=\frac{-2}{3} C(\mu)\xi_{1}+\frac{2(1+\omega)}{3} C(\mu)\xi_{2}},
\\ {N_{1}=\displaystyle\frac{-1}{\sqrt{3}} C(\mu)\xi_{1}+\frac{\omega-1}{\sqrt{3}} C(\mu)\xi_{2}, 
\qquad K_{1}=\displaystyle\frac{-1}{\sqrt{3}} C(\mu)\xi_{1}+\frac{\omega+1}{\sqrt{3}} C(\mu)\xi_{2}}, 
\\ {N_{2}=0, \qquad K_{2}=0}, 
\\ {N_{3}=\displaystyle0+\frac{2}{\sqrt{3}} C(\mu)\xi_{2}, \qquad K_{3}=0+\frac{-2}{\sqrt{3}} C(\mu)\xi_{2}}, 
\\ {N_{4}=\displaystyle\frac{-2}{\sqrt{3}} C(\mu)\xi_{1}+0, \qquad K_{4}=\frac{2}{\sqrt{3}} C(\mu)\xi_{1}+0}.
\end{array}
\end{equation}

\subsection{QCD sum rule and Large $N_c$}  

Isgur-Wise functions and the mass parameters should be calculated by 
nonperturbative methods.  In this work, we make use of results from QCD 
sum rule \cite{huang2005note,dai1996qcd} and large $N_c$ methods 
\cite{jenkins1993baryon,lee1998analysis,Du:2011nj,Du:2013zba}.  
\subsubsection{QCD sum rule}
Within HQET, the QCD sum rule method gives the following results 
\cite{dai1996qcd,huang2005note},  
\begin{equation}
\label{qcdsr}
\begin{array}{lll}
\displaystyle\xi(\omega)     &=     &1-\rho^{2}(\omega-1), \quad \rho^{2}=1.35 \pm 0.12, 
\\[3mm]
\displaystyle\chi(\omega)   &\simeq&O\left(10^{-2}\right), \\[3mm] 
\displaystyle\bar{\Lambda}&\simeq&0.79 \pm 0.05 \mathrm{GeV}\,.
\end{array}
\end{equation}  
\subsubsection{Large $N_c$}
In the large $N_c$ limit, the leading Isgur-Wise function $\xi(\omega)$ and the mass 
parameter $\bar{\Lambda}$ are given as \cite{jenkins1993baryon}
\begin{equation}
\xi(\omega)=0.99 \exp [-1.3(\omega-1)], ~~~ 
\bar{\Lambda}\simeq 0.87 ~ \mathrm{GeV}.  
\end{equation}
This $\xi$ is actually a reallization of $\delta$ function 
\cite{lee1998analysis}.   Ref.\cite{lee1998analysis} further showed 
that $\chi(\omega)=0$ in the large $N_c$ limit.  

In the large $N_c$ limit, Isgur-Wise functions $\xi_1$and $\xi_2$ can 
be expressed by $\xi(\omega)$ \cite{Chow:1994ni,Chow:1996xf}, 
\begin{equation}
\xi_{1}(\omega)=\xi(\omega), \quad \xi_{2}(\omega)=\frac{\xi(\omega)}{1+\omega}.
\end{equation} 
To the order of $1/m_Q$, sub-leading Isgur-Wise functions can also be 
written as $\xi(\omega)$ \cite{Du:2011nj,Du:2013zba}.  We finally 
obtain the form factors as 
\begin{equation}
\begin{aligned}
F_{1}^{\prime}
    &=-C(\mu)\frac{1}{3} \xi(\omega)-\frac{1}{3} C(\mu)\xi(\omega)\left[\frac{\bar{\Omega}}{2 m_{c}}+\frac{\bar{\Omega}}{2 m_{b}}\right], 
\\ 
F_{2}^{\prime}
    &=C(\mu)\frac{4 \xi(\omega)}{3(1+\omega)}+C(\mu)\frac{\xi(\omega)}{3(1+\omega)}\left[-\frac{\bar{\Omega}}{m_{c}}+\frac{2 \bar{\Omega}}{m_{b}}\right],
\\ 
F_{3}^{\prime}
    &=C(\mu)\frac{4 \xi(\omega)}{3(1+\omega)}+C(\mu)\frac{\xi(\omega)}{3(1+\omega)}\left[\frac{2 \bar{\Omega}}{m_{c}}-\frac{\bar{\Omega}}{m_{b}}\right],
\\
G_{1}^{\prime}
    &=-C(\mu)\frac{1}{3} \xi(\omega)+C(\mu)\frac{1}{3} \xi(\omega)\left[\frac{\bar{\Omega}}{2 m_{c}}+\frac{\bar{\Omega}}{2 m_{b}}\right]\left(\frac{1-\omega}{1+\omega}\right), 
\\ 
G_{2}^{\prime}
    &=C(\mu)\frac{\bar{\Omega}}{3 m_{c}}\left(\frac{1}{1+\omega}\right) \xi(\omega),
\\ G_{3}^{\prime}
    &=-C(\mu)\frac{\bar{\Omega}}{3 m_{b}}\left(\frac{1}{1+\omega}\right) \xi(\omega),
\\N_{1}
    &=C(\mu)\frac{-2 \xi(\omega)}{\sqrt{3}(1+\omega)}+C(\mu)\frac{-\xi(\omega)}{\sqrt{3}(1+\omega)}\left[\frac{\bar{\Omega}}{m_{c}}+\frac{\bar{\Omega}}{m_{b}}\right], 
    \\
    K_{1}&=0, \quad
N_{2}
    =0, \quad K_{2}=C(\mu)\frac{2}{\sqrt{3}} \xi(\omega)\frac{\bar{\Omega}}{m_{c}}\left(\frac{1}{1+\omega}\right)^{2}, 
\\ 
N_{3}
    &=C(\mu)\frac{2 \xi(\omega)}{\sqrt{3}(1+\omega)}+C(\mu)\frac{\xi(\omega)}{\sqrt{3}(1+\omega)}\left[\frac{\bar{\Omega}}{m_{c}}+\frac{\bar{\Omega}}{m_{b}}\right],
\end{aligned}
\end{equation}
where $\bar{\Omega}=m_{\Omega_{Q}}-m_{Q}$.

\section{Numerical results}

Numerical results for 
$\Lambda_{b}\rightarrow\Lambda_{c}l\bar{\nu}_l$ and 
$\Omega_{b}\rightarrow\Omega_{c}^{(*)}l\bar{\nu}_l$ ($l=e,\mu,\tau$) 
can be obtained now.  In the calculation it takes $m_{\Lambda_b}=5.62$ 
GeV, $m_{\Lambda_c}=2.23$ GeV, $m_{\Omega_b}=6.07$ GeV, 
$m_{\Omega_c}=2.70$ GeV, $m_{\Omega_c^*}=2.77$ GeV, 
$\left|V_{cb}\right|=0.04$ and $G_F=1.166\times 10^{-5}$ GeV$^{-2}$ 
\cite{beringer2020particle}.  And $m_c=1.44$ GeV, $m_b=4.83$ GeV, 
$\mu=0.47$ GeV \cite{manohar2007heavy,Neubert:1993mb}.  $\omega$ is in 
the range 
$1\leq\omega\leq\dfrac{m_{\Lambda_b}^2+m_{\Lambda_c}^2-m_\tau^2}{2m_{\Lambda_b}m_{\Lambda_c}}$ 
for the $\Lambda_b\to\Lambda_c$ decay and 
$1\leq\omega\leq\dfrac{m_{\Omega_b}^2+m_{\Omega_c^{(*)}}^2-m_\tau^2}{2m_{\Omega_b}m_{\Omega_c^{(*)}}}$ 
for $\Omega_b\to\Omega_c^{(*)}$.  

Tauonic decay distributions are plotted in Figs.\,1-7.  Fig.\,1 
presents the $\Lambda_b\to\Lambda_c\tau\bar{\nu}_\tau$ differential 
decay rate, both QCD sum rule and large $N_c$ results are given for 
comparison, with the uncertainty of the QCD sum rule considered.  
The two results are close to each other, especially in the low recoil 
region.  In Figs.\,2 and 3, we display the $\omega$ dependence of 
$\Lambda_{b}\rightarrow\Lambda_{c}\tau\bar{\nu}_\tau$ partial 
differential rates $T, L, t$ and the total differential rate.  The 
transverse rate $T$ dominates in the low recoil region while the 
longitudinal rate $L$ dominates in the large recoil region.  Fig.\,2 is 
for the QCD sum rule method.  And Fig.\,3 is that from the large $N_c$ 
method.  Figs.\,4-7 show the corresponding plots for 
$\Omega_b\to\Omega_c^{(*)}\tau\bar{\nu}_\tau$ decays for the large 
$N_c$ limit.  For the partial decay distribution of 
$\Omega_{b}\rightarrow\Omega_{c}^{*}\tau\bar{\nu}_{\tau}$ (Fig.\,7), 
what should be discussed is that the $t_{+}$ channel is almost $0$, and 
the $L_{\pm}$ channels are almost the same.  As for the tauonic decay, 
time-components should be considered specifically, because they are absent 
in the massless charged lepton case.  In the $\Lambda_{b}$ case, 
time-component is still small.  However, in the $\Omega_b\to\Omega_c$ 
case, time-component gets comparatively larger. In the 
$\Omega_b\to\Omega_c^*$ case, time-component gets to be even much larger 
and begins to dominate in the large recoil region. 

The decay rates are obtained by $\omega$ integration.  For the 
$\Lambda_{b}\to\Lambda_{c}\tau\bar{\nu}_{\tau}$ decay, we obtain the 
total decay rate, the branching ratio, and the R-ratio in the following  
from the QCD sum rule, 
\begin{equation}\displaystyle
\label{qcdsr2}
\begin{array}{lll}
\displaystyle \Gamma \left(\Lambda_{b}\to\Lambda_{c}\tau\bar{\nu}_{\tau}\right)&=&1.16\pm0.05~(\rho^2)\pm0.004~(\bar{\Lambda})\times10^{-14} \ \mathrm{GeV}\,, \\[3mm] 
\displaystyle {\rm Br} \left(\Lambda_{b}\to\Lambda_{c}\tau\bar{\nu}_{\tau}\right)&=&\displaystyle (2.59\pm 0.09)\%\times\left(\frac{\tau\left(\Lambda_{b}\right)}{1.47\times 10^{-12} \mathrm{sec}}\right)\,, \\[3mm]
{\displaystyle} R\left(\Lambda_{c}\right)&=&\displaystyle\frac{\Gamma\left(\Lambda_{b}\to\Lambda_{c}\tau\bar{\nu}_{\tau}\right)}{\Gamma\left(\Lambda_{b}\to\Lambda_{c}\mu\bar{\nu}_{\mu}\right)}=(33.1\pm 1.4)\% \,, 
\end{array}
\end{equation}  
where uncertainties are due to the error of QCD sum rules in Eq. (\ref{qcdsr}).  For the 
large $N_c$ case, 
\begin{equation}\label{qcdsr3}
\begin{array}{lll}
\Gamma\left(\Lambda_{b}\to\Lambda_{c}\tau\bar{\nu}_{\tau}\right)&=&1.22\times 10^{-14} \ \mathrm{GeV}\,,\\[3mm] 
{\rm Br}\left(\Lambda_{b}\to\Lambda_{c}\tau\bar{\nu}_{\tau}\right)&=&2.73\%\   \displaystyle\times\left(\frac{\tau\left(\Lambda_{b}\right)}{1.47 \times 10^{-12} \mathrm{sec}}\right)\,,\\[3mm] 
R\left(\Lambda_{c}\right)&=&\displaystyle\frac{\Gamma\left(\Lambda_{b}\to\Lambda_{c}\tau\bar{\nu}_{\tau}\right)}{\Gamma\left(\Lambda_{b}\to\Lambda_{c}\mu\bar{\nu}_{\mu}\right)}=29.2\%\,.
\end{array}   
\end{equation} 
The error of the large $N_c$ result is estimated to be $1/N_c\sim 30\%$ 
in general.  However, the uncertainty of $R$ which is what we are 
really interested in, is supposed to be smaller because of the cancellation in the 
ratios \cite{lee1998analysis}.  Thus, the uncertainty in 
$R\left(\Lambda_{c}\right)$ is estimated as small as $\sim 10\%$.  

Table I lists results for the 
$\Lambda_b\to\Lambda_c \ell{\bar{\nu}}_\ell$ semileptonic decay.  
Experimental data \cite{beringer2020particle} and results from the 
quark model \cite{gutsche2015semileptonic}, HQET \cite{bernlochner2018new}, and  lattice QCD 
\cite{Detmold:2015aaa} are also listed for comparison.  
From the table one can see that the large $N_c$ result is somewhat 
larger than other theoretical results, as far as central values are 
concerned.  The QCD sum rule result is very close to that of the quark 
model.  Nevertheless within the $2\sigma$ uncertainty, all the results 
are still consistent with each other.   

Via the same procedure, the result of the 
$\Omega_b\rightarrow\Omega_c^{(*)}\tau\bar{\nu}_{\tau}$ decay is 
obtained by using the large $N_c$ method in the following, 
 \begin{equation}\label{qcdsr4}
 \begin{aligned}
 \Gamma\left(\Omega_{b} \rightarrow \Omega_{c} \tau \bar{\nu}_{\tau}\right)&=4.83\times10^{-15} \ \mathrm{GeV},
 \\
{\rm Br}\left(\Omega_{b} \rightarrow \Omega_{c} \tau \bar{\nu}_{\tau}\right)&= 1.21\%\   \displaystyle\times\left(\frac{\tau\left(\Omega_{b}\right)}{1.65 \times 10^{-12} \mathrm{sec}}\right)\,,
\\ 
\displaystyle R\left(\Omega_{c}\right)&=\frac{\Gamma\left(\Omega_{b} \rightarrow \Omega_{c} \tau \bar{\nu}_{\tau}\right)}{\Gamma\left(\Omega_{b} \rightarrow \Omega_{c} \mu \bar{\nu}_{\mu}\right)}=30.4\%.
  \end{aligned}
  \end{equation}  
And 
 \begin{equation}\label{acdsr5}
 \begin{aligned}
 \displaystyle
 \Gamma\left(\Omega_{b} \rightarrow \Omega_{c}^{*} \tau \bar{\nu}_{\tau}\right)&=1.27\times10^{-14}  \ \mathrm{GeV},
\\
{\rm Br}\left(\Omega_{b} \rightarrow \Omega_{c}^{*} \tau \bar{\nu}_{\tau}\right)&=3.18\%\   \displaystyle\times\left(\frac{\tau\left(\Omega_{b}\right)}{1.65 \times 10^{-12} \mathrm{sec}}\right)\,, 
\\
 \displaystyle R\left(\Omega_c^*\right)&=\frac{\Gamma\left(\Omega_{b} \rightarrow \Omega_{c}^{*}\tau\bar{\nu}_{\tau}\right)}{\Gamma\left(\Omega_{b} \rightarrow \Omega_{c}^{*} \mu \bar{\nu}_{\mu}\right)}=33.2\%.
 \end{aligned}
 \end{equation}
Like in the $\Lambda_b$ decay, the $1/N_c$ uncertainty for 
$R(\Omega_c^{(*)})$ is expected to be $\sim 10\%$. 

\begin{table}[bth!]
\caption{Results for the $\Lambda_b\to\Lambda_c \ell{\bar{\nu}}_\ell$ decay.} 
\label{compdecaywidth} 
\begin{threeparttable} 
\begin{ruledtabular}
\begin{tabular}{ccccccccc}
& Sum rule & Large $N_c$ &  Ref.\cite{gutsche2015semileptonic} \tnote{1} &  Ref.\cite{Detmold:2015aaa}\tnote{2} & Ref.\cite{bernlochner2018new}\tnote{3} &  Experiment\cite{beringer2020particle}
\\
\hline
$\Lambda_b\to\Lambda_c (e,\mu) {\bar{\nu}}_{(e,\mu)}$
\\
$    \Gamma\times10^{14}$     &       $3.50\pm0.30$      &     $4.2\pm0.4$      &     3.13  &    $2.26\pm0.14$        &         &    
\\
$Br$ (\%)                                   
&    $7.83\pm0.60$            & $9.3 \pm0.9$                    &      7.0      &      $5.06\pm0.32$     &            &  $6.2^{+1.4}_{-1}$           
\\
\hline
$\Lambda_b\to \Lambda_c \tau \bar{\nu}_{\tau}$
\\
$\Gamma \times10^{14}$     
&       $ 1.16\pm 0.05 $      & $1.2\pm0.1 $             &0.91         &    $0.753\pm0.033$     &          &                                   
\\
$Br$ (\%)                                   
&       $ 2.59\pm0.09 $       & $2.7\pm0.3$              &2.0          &   $1.68\pm0.07$      &  
\\
\hline

$R\left(\Lambda_{c}\right)$
&     $0.33\pm0.01$                &    $0.29\pm0.03$    &     $0.29$    &     $0.3328\pm0.0102$   &     $0.324\pm0.004$
\end{tabular}
\end{ruledtabular}
\begin{tablenotes}    
        \footnotesize               
        \item[1] \tiny Covariant confined quark model.    
        \item[2]  Lattice QCD .
        \item[3] HQET to $\mathcal{O}\left(\Lambda_{\mathrm{QCD}}^{2} / m_{c}^{2}\right)$, form factors are determined by  fitting to LHCb and lattice QCD data. 
      \end{tablenotes}            
    \end{threeparttable}  
\end{table}

\begin{figure}
    \centering
    \includegraphics[height=4.5cm,width=9.5cm]{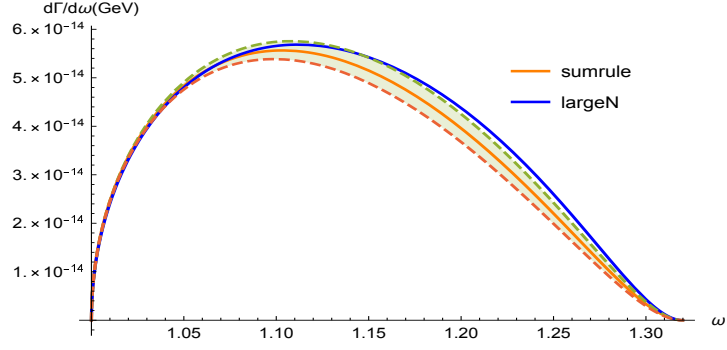}
    \caption{The differential decay rate of $\Lambda_{b}\to\Lambda_{c}\tau\bar{\nu}_{\tau}$ from QCD sum rule and large $N_c$ methods.  $1\leq\omega\leq1.31  $.}
\end{figure}


\begin{figure}
    \centering
    \includegraphics[height=4.5cm,width=9.5cm]{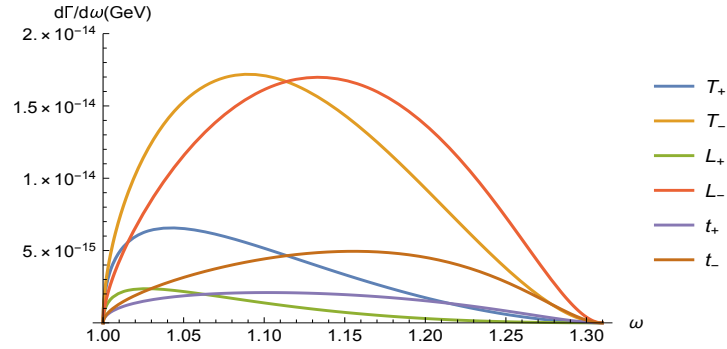}
    \caption{Partial decay distributions in various helicities of $\Lambda_{b}\to\Lambda_{c}\tau\bar{\nu}_{\tau}$ from the QCD sum rule.  $T,~L,~t$ are defined in Eq. (\ref{lambda}), $1\leq\omega\leq1.31  $.}
\end{figure}

\begin{figure}
    \centering
    \includegraphics[height=4.5cm,width=9.5cm]{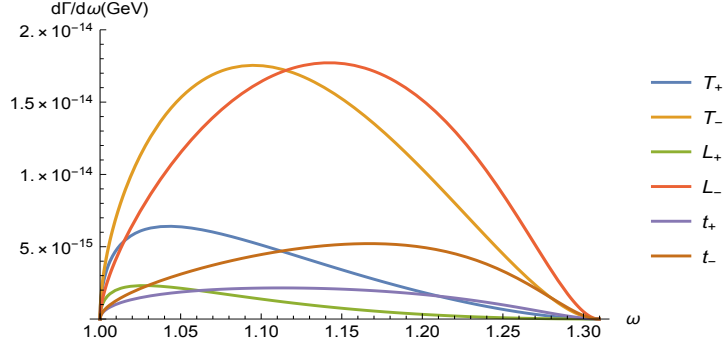}
    \caption{Partial decay distributions in various helicities of $\Lambda_{b}\to\Lambda_{c}\tau\bar{\nu}_{\tau}$ in the Large $N_c \ QCD$. $1\leq\omega\leq1.31$}
\end{figure}

\begin{figure}
    \centering
    \includegraphics[height=4.5cm,width=9.5cm]{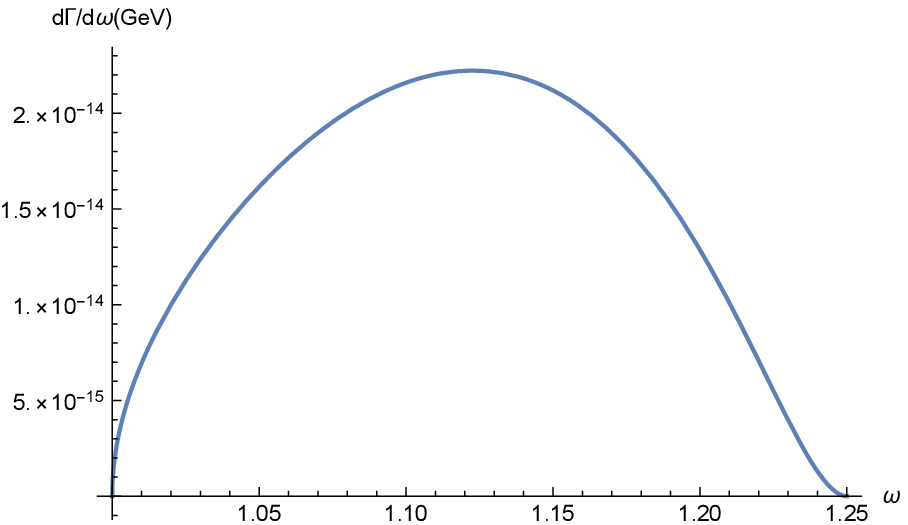}
    \caption{The differential decay rate of $\Omega_{b}\rightarrow \Omega_{c}\tau\bar{\nu}_{\tau}$, $1\leq\omega\leq1.25 $}
\end{figure}
\begin{figure}
    \centering
    \includegraphics[height=4.5cm,width=9.5cm]{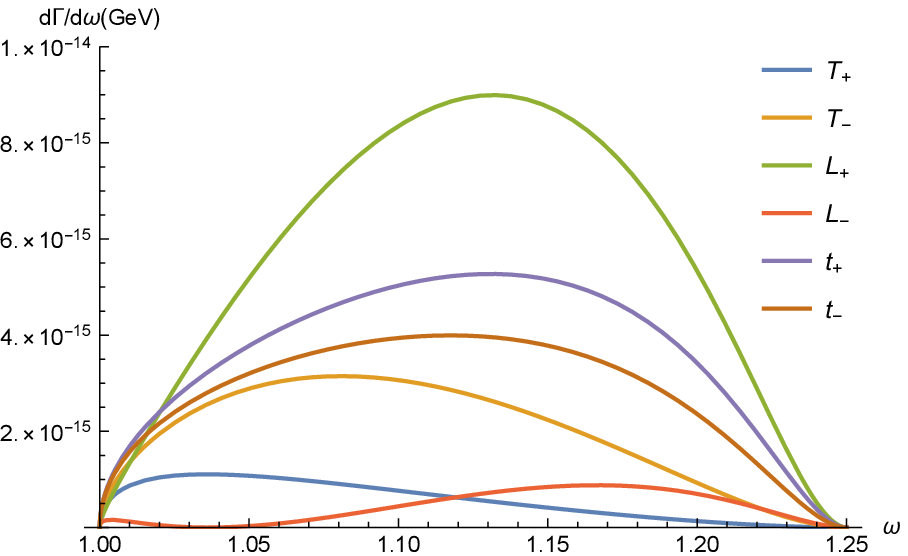}
    
    \caption{Partial decay distributions in various helicities of $\Omega_{b}\to\Omega_{c}\tau\bar{\nu}_{\tau}$. T, L, t stand for $\frac{\mathrm{d} \Gamma^{\prime}_{T_{\pm}}}{\mathrm{d} \omega}\frac{\mathrm{d} \Gamma^{\prime}_{L_{\pm}}}{\mathrm{d} \omega},\frac{\mathrm{d} \Gamma^{\prime}_{t_{\pm}}}{\mathrm{d} \omega}$  which are defined in Eq. (\ref{omega}), $1\leq\omega\leq1.25 $ }
\end{figure}
\begin{figure}
    \centering
    \includegraphics[height=4.5cm,width=9.5cm]{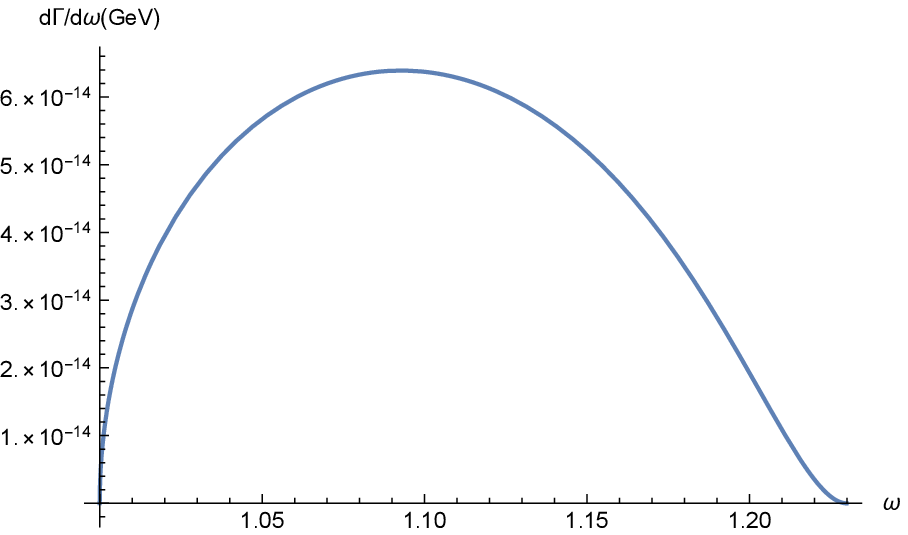}
    \caption{The differential decay rate of $\Omega_{b}\to\Omega_{c}^{*}\tau\bar{\nu}_{\tau}$, $1\leq\omega\leq1.23$}
\end{figure}
\begin{figure}
    \centering
    \includegraphics[height=4.5cm,width=9.5cm]{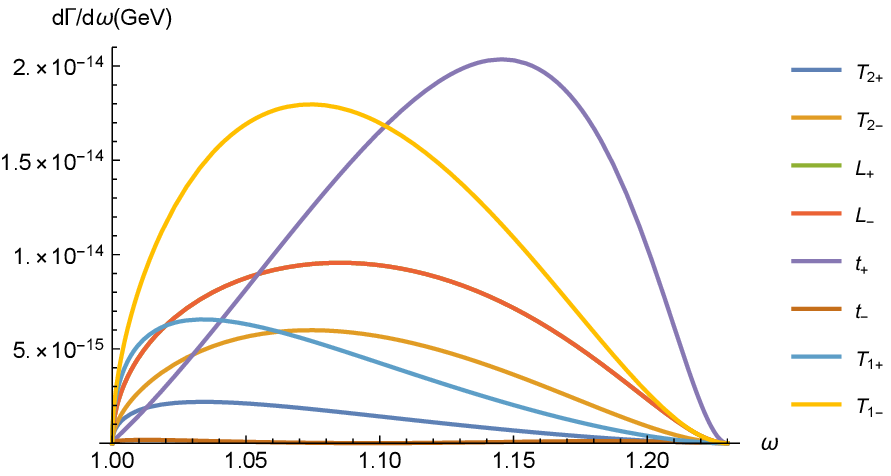}
    \caption{Partial decay distributions in various helicities of $\Omega_{b}\to\Omega_{c}^{*}\tau\bar{\nu}_{\tau}$, T, L, t stand for $\frac{\mathrm{d} \Gamma^{\prime\prime}_{T_{\pm}}}{\mathrm{d} \omega}\frac{\mathrm{d} \Gamma^{\prime\prime}_{L_{\pm}}}{\mathrm{d} \omega},\frac{\mathrm{d} \Gamma^{\prime\prime}_{t_{\pm}}}{\mathrm{d} \omega}$ which are defined in Eq. (\ref{omega*}), $1\leq\omega\leq1.23$.}
\end{figure}

\section{Summary}

In this paper, $\Lambda_b\rightarrow\Lambda_c\tau\bar{\nu}_\tau$ and 
$\Omega_b\rightarrow\Omega_{c}^{(*)}\tau\bar{\nu}_{\tau}$ semileptonic 
decays have been calculated within the Standard model systematically.  
In the analysis with lepton-mass effects considered, helicity amplitudes 
have been given, form factors are expanded in HQET to the order of 
$\Lambda_{\rm QCD}/m_b$ and $\Lambda_{\rm QCD}/m_c$, the Isgur-Wise 
functions obtained by QCD sum rule and large $N_c$ methods have been 
applied in the calculation.  We have obtained decay rates, decay 
distributions, and R-ratios for 
$\Lambda_b\rightarrow\Lambda_c\tau\bar{\nu}_\tau$ and 
$\Omega_b\rightarrow\Omega_{c}^{(*)}\tau\bar{\nu}_{\tau}$ decays.  The 
R-ratio $R\left(\Lambda_{c}\right)\simeq (33\pm 1)\%$ (QCD sum rule), 
$R\left(\Lambda_{c}\right)\simeq (29\pm3)\%$ (large $N_c$ QCD), 
$R\left(\Omega_{c}\right)\simeq 30.4\%$ (large $N_c$ QCD), and 
$R\left(\Omega_c^*\right)=33.2\%$ (large $N_c$ QCD) with an estimated 10\% uncertainty for the large $N_c$.  These results 
will be checked by experiments in the near future, such as LHCb, to see 
if there is any new physics in these decays.

\begin{acknowledgments}

We are very grateful to J\"{u}rgen G. K\"{o}rner for valuable comments.  
We acknowledge support from the National Natural Science Foundation of 
China (No. 11875306).
\end{acknowledgments}

\bibliographystyle{JHEP}
\bibliography{dust}

\end{document}